%% file: main.tex
\documentclass{IEEEcsmag}

\usepackage[colorlinks,urlcolor=blue,linkcolor=blue,citecolor=blue]{hyperref}
\usepackage{tabu}                      
\usepackage{booktabs}                  
\usepackage{multirow}
\usepackage{upmath}
\usepackage{xcolor}

\jvol{XX}
\jnum{XX}
\paper{8}
\jmonth{August}
\jname{IT Professional}
\pubyear{2022}

\begin{document}

\sptitle{Department: Head}
\editor{Editor: Name, xxxx@email}

\newcommand{\xml}[1]{{\color{cyan}#1}}
\newcommand{\techName}[1]{\textit{DiffSeer}}
\newcommand{\yong}[1]{\textcolor{brown}{[Yong: #1]}}
\newcommand{\ff}[1]{\textcolor{purple}{[FF: #1]}}
\newcommand{\wen}[1]{\textcolor{pink}{[Wen: #1]}}
\newcommand{\wmx}[1]{\textcolor{orange}{[WMX: #1]}}
\newcommand{\mod}[1]{\textcolor{black}{#1}}
\newcommand{\newmod}[1]{\textcolor{black}{#1}}

\title{\techName{}: Difference-based Dynamic Weighted Graph Visualization}


\author{Xiaolin Wen}
\affil{Sichuan University, Chengdu, China and Singapore Management University, Singapore}

\author{Yong Wang}
\affil{Singapore Management University, Singapore}

\author{Meixuan Wu, Fengjie Wang}
\affil{Sichuan University, Chengdu, China}

\author{Xuanwu Yue}
\affil{Sinovation Ventures AI Institute, Beijing, China and Baiont Technology, Nanjing, China}

\author{Qiaomu Shen, Yuxin Ma}
\affil{Southern University of Science and Technology, Shenzhen, China}

\author{Min Zhu}
\affil{Sichuan University, Chengdu, China}

\markboth{Department Head}{Paper title}

\begin{abstract}
\mod{Existing dynamic weighted graph visualization
approaches
rely on users' mental comparison to perceive temporal evolution of dynamic weighted graphs, hindering users from effectively analyzing changes across multiple timeslices.}
We propose \techName{}, a novel approach for dynamic weighted graph visualization by explicitly visualizing the \textit{differences} of graph structures (e.g., edge weight differences) between adjacent timeslices.
Specifically, we present a novel nested matrix design that overviews the graph structure differences over a time period as well as shows graph structure details in the timeslices of user interest.
By collectively considering the overall temporal evolution and structure details in each timeslice, an optimization-based node reordering strategy is developed to group nodes with similar evolution patterns and highlight interesting graph structure details in each timeslice.
We conducted two case studies on real-world graph datasets and in-depth interviews with 12 target users to evaluate \techName{}. 
The results demonstrate its effectiveness in visualizing dynamic weighted graphs.
\end{abstract}

\maketitle


\input{source/1-intro}

\input{source/2-relatedwork}
\input{source/3-background}
\input{source/4-diffSeer}

\input{source/5-usecases}

\input{source/6-expertinterview}
\input{source/7-discussion}
\input{source/8-conclusion}


\bibliographystyle{abbrv-doi}

\bibliography{main}
\end{document}

%% file: source/1-intro.tex
%
%

\chapterinitial{Dynamic weighted graphs} model the temporal evolution of detailed relationships between entities in various applications such as social networks, financial networks, or communication networks.
To analyze such dynamic weighted graphs,
a large number of dynamic graph visualization techniques have been proposed, where the essential research question is to investigate how to visualize \textit{the temporal changes} of dynamic graph structures effectively~\cite{Ahn2013task,yi2010timematrix}.
Most existing dynamic graph visualization approaches focus on displaying graph structures (i.e., nodes and edges) along the time via either animated diagrams or a series of static charts (e.g., small multiples)~\cite{beck2017taxonomy}.
\newmod{It is non-trivial to compare the differences between weighted graphs~\cite{alper2013weighted}.}
To explore the temporal evolution patterns of dynamic weighted graphs,
users need to \textit{mentally compare} the differences between multiple adjacent timeslices simultaneously, which is more challenging.

\begin{figure}
  \centering
  \includegraphics[width=0.95\linewidth]{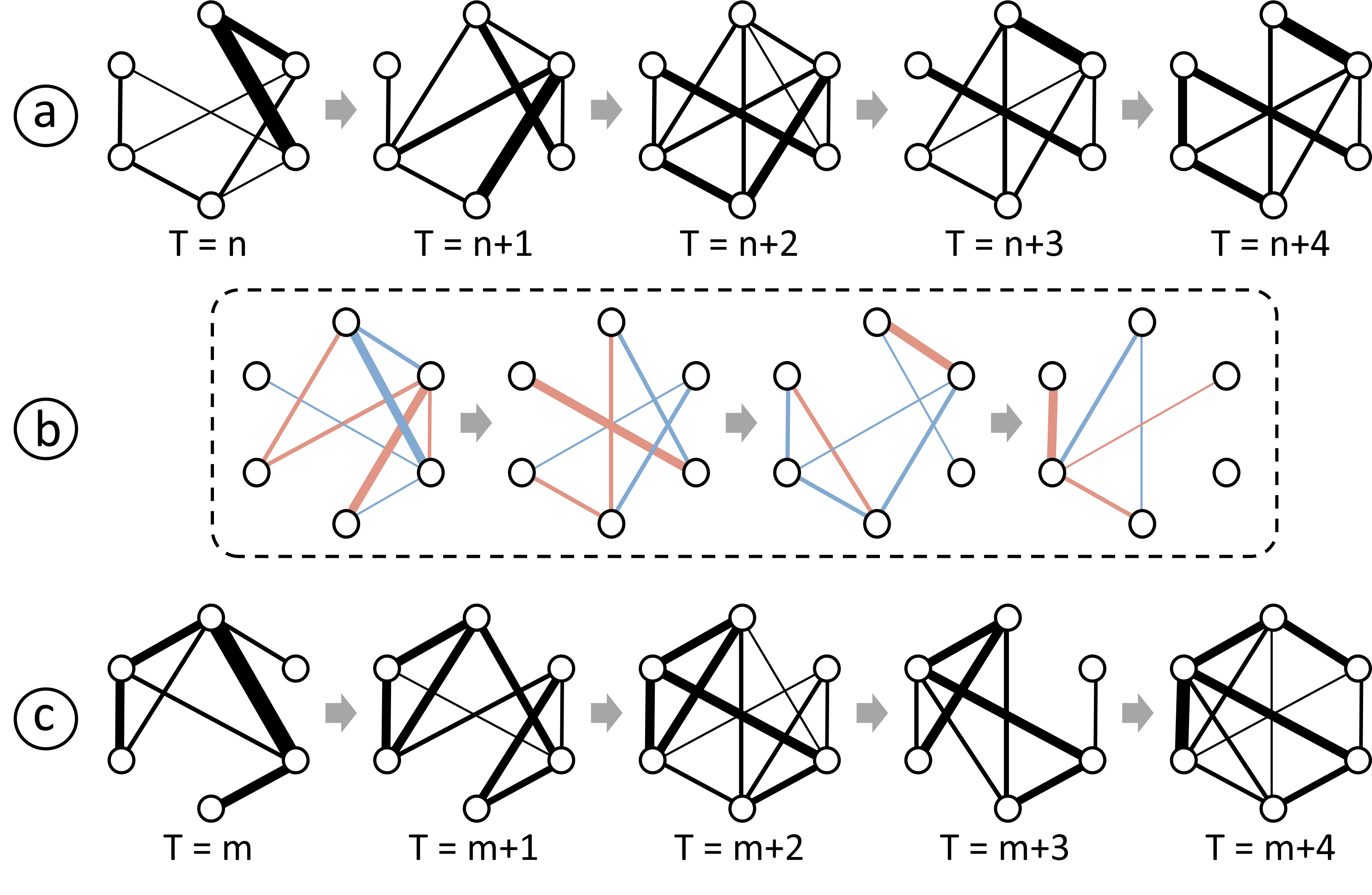}
    \setlength{\abovecaptionskip}{0cm} 
    \setlength{\belowcaptionskip}{-2.0cm}
  \caption{Illustration of the importance of differences in dynamic weighted graph visualization.\mod{ (a) and (c) show two different dynamic weighted graphs.
  (b) shows the graph differences between adjacent timeslices of both (a) and (c), where the edge width encodes the edge weight changes and the color represents the trend of changes (i.e., red for an increase and blue for a decrease). (b) provides a direct perception of temporal changes in dynamic weighted graphs.}}
          \label{fig:intro}
\end{figure}

Different from prior studies for dynamic graph visualization, we aim to
achieve effective dynamic weighted graph visualization across a period of time from a new perspective:
\textit{explicitly visualizing the \textit{differences} between adjacent timeslices}.
Such a new perspective has clear advantages.
As shown in Figure~\ref{fig:intro}a, it's hard to quickly identify how the graph structures (e.g., edge weights) are exactly changing between two adjacent timeslices through mental comparison.
\mod{
When explicitly encoding the differences of edge weight by the width and using red and blue to represent the positive and negative changes (Figure~\ref{fig:intro}b), the temporal changes of dynamic weighted graphs can be seen directly and do not rely on the mental comparison.
Figure~\ref{fig:intro}c shows another dynamic weighted graph that is different from Figure~\ref{fig:intro}a, but they have the same difference sequence (Figure~\ref{fig:intro}b).
}
\mod{
However, it is non-trivial to leverage differences for dynamic graph visualizations.
First, graph differences can refer to different aspects of graphs such as the changes of edge weights. How to visualize such graph differences in a universal way needs further exploration.
Second, the temporal evolution of dynamic weighted graphs relies on understanding multiple graph differences of continuous adjacent timeslices.
It remains unclear regarding how to visualize such multiple graph differences over time and highlight the temporal patterns like reoccurring and outliers~\cite{van2015reducing}.}
Third, despite the importance of graph differences, graph structures themselves are also useful for a comprehensive interpretation of dynamic weighted graphs.
\mod{
It is challenging to inform users of both graph differences and original dynamic weighted graph structures
effectively.
}
%
%

\begin{figure*}
 \includegraphics[width=\linewidth]{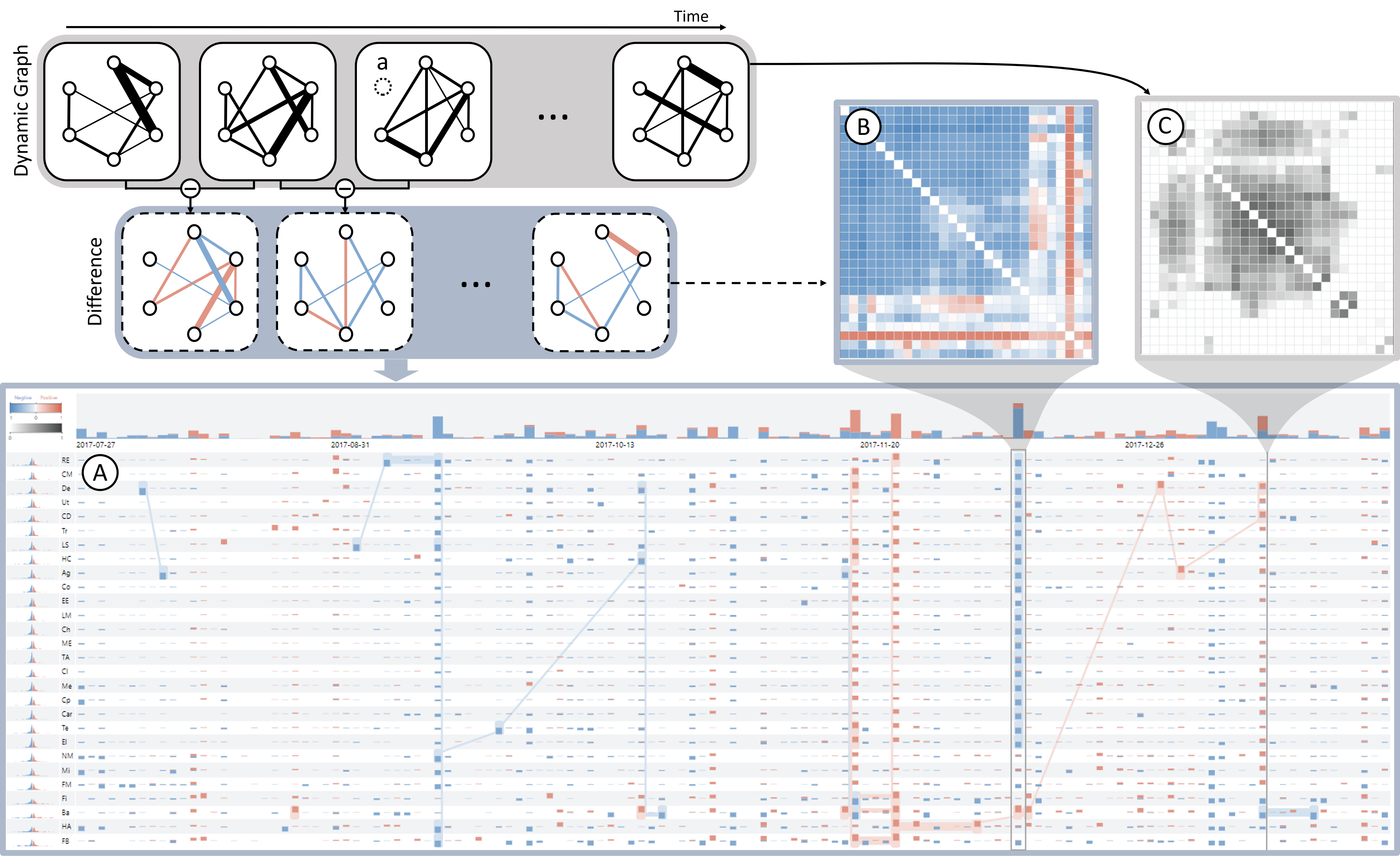}
 \centering
  \setlength{\abovecaptionskip}{-0.5cm}
  \caption{Overview of \techName{}: We focus on explicitly visualizing the differences between adjacent timeslices to support the analysis of the dynamic weighted graph evolution over a long time. Specifically, we proposed a \textit{nested matrix design}, including (A) an \textit{overview matrix} to provide a visual summary of differences and two types (B, C) of \textit{detail matrices} to enable interactive inspection of graph details on demand. An optimization-based \textit{node reordering strategy} is incorporated in the nested matrix design to group together nodes with similar evolution patterns and highlight interesting graph structure details in each timeslice.}
\label{fig:teaser}
\end{figure*}

%
%
In this work, 
we propose \techName{}, a novel difference-based approach for dynamic weighted graph visualization
(Figure~\ref{fig:teaser}).
It can effectively inform users of the overall temporal evolution of dynamic weighted graphs and show the graph structure details in individual timeslices.
Specifically, we present a novel \textit{nested matrix design}
that overviews the graph differences over a period of time and enables interactive inspection of graph details on demand.
Given the essential information of dynamic weighted graphs is edge weights~\cite{bach2014visualizing}, we leverage edge weight differences to represent graph differences.
A new \textit{node reordering strategy} is proposed to
group nodes in the nested matrix design, which can manifest the overall temporal evolution patterns and interesting graph structure details of individual timeslices in a balanced manner.
A \textit{difference mask} is integrated into the nested matrix design to enhance the perception of interesting temporal evolution patterns.
We conduct two case studies on real-world dynamic graph datasets and in-depth interviews with 12 expert users.
The results demonstrate the effectiveness and usability of \techName{} in informing viewers of the temporal evolution of dynamic weighted graphs.

In summary, our main contributions include:
\begin{itemize}
    \item We propose \techName{}, a novel difference-based dynamic weighted graph visualization approach, for exploring the dynamic graph evolution interactively, where
    a novel nested matrix design is presented to show the temporal evolution of dynamic graphs with multiple levels of details.
    \item \mod{We conduct two case studies on real-world datasets and user interviews with 12 target users to demonstrate the effectiveness and usability of \techName{}.}
\end{itemize}




%% file: source/2-relatedwork.tex
\section{Related Work}

Dynamic graph visualization has been extensively studied in the past decades.
The major dynamic graph visualization methods include animation-based and timeline-based approaches~\cite{beck2017taxonomy}.
Animation-based approaches show the temporal evolution through animated transitions~\cite{rufiange2014animatrix}.
Timeline-based approaches visualize dynamic graphs through a time-to-space mapping, where the graph structure of each timeslice is shown along a timeline~\cite{bach2015small,bach2014visualizing}.
Most of these approaches focus on displaying the graph structures in each timeslice directly.
To identify the temporal evolution patterns of dynamic graphs, users need to mentally compare the changes or differences between adjacent timeslices in the animated graphs or between a sequence of static graphs~\cite{archambault2010animation}.
\mod{
For dynamic weighted graphs, such a process relies on users’ mental memory due to the complex weight changes, hindering users from effectively and quickly analyzing the temporal changes.}

\mod{
A few prior studies have also investigated highlighting the removed or inserted edges and nodes between two adjacent timeslices in small multiples~\cite{archambault2009structural,rufiange2013diffani} and animated transitions~\cite{bach2013graphdiaries,crnovrsanin2020staged}.
Archambault et al.~\cite{archambault2010difference} further performed a user study on such methods and showed that highlighting changing edges is helpful.}
\mod{
However, these studies focus on using color to indicate whether an edge occurred or not
and are unable to reveal change details
in dynamic weighted graphs such as edge weight~\cite{bach2014visualizing}.
\newmod{
For dynamic weighted graph, users need to understand both the edge weight differences and the original graph structures.}
\newmod{
Previous studies have never explored leveraging detailed edge weight differences to display the evolution of dynamic weighted graphs.}
}



\newmod{
Different from prior studies, \techName{} is a novel difference-based visualization approach for dynamic weighted graphs. It explicitly visualizes the edge weight differences over time, and enables systematic exploration of the temporal evolution of dynamic weighted graphs with multiple levels of details.}

%% file: source/3-background.tex
\section{Background}\label{sec:tech}



In this paper, we focus on the undirected dynamic weighted graph and give concrete definitions as follows.
A \textit{dynamic weighted graph} $ \Gamma $ can be regarded as a sequence of graph snapshots $G$ in each timeslice:
\begin{equation}
    \Gamma = \left \{ G_{1},G_{2},...,G_{T}  \right \}, 
\end{equation}
where each graph snapshot $G_{i} = (V_{i},E_{i})$
has a different node set $ V_{i} =\left \{v_{1},v_{2},...,v_{N}\right \} $ and a different edge set $ E_{i}=\left \{e_{1},e_{2},...,e_{M}\right \} $ in each timeslice. 
Each edge $ e_m $ is defined as follows: 
\begin{equation}
    e_m = (u,w,v) \in  E_{i} \subseteq V_{i} \times R ^ {+} \times V_{i},
\end{equation}
where $u, v \in V_{i}$ are two nodes linked by $ e_m $, $w$ denotes the weights of edges in the graph.

\mod{
We define the node set $V$ as all the nodes that appear at least once throughout the whole time range, and define the \textit{graph difference} between two adjacent timeslices of a dynamic weighted graph as follows:}
\begin{equation}\label{diff}
     Diff_{i} = (V,D_{i}),
\end{equation}
where 
$ D_{i} $ is the set of edges with weight changes in the i-th timeslice.
${Diff}_{i}$ is intrinsically a graph as well.
Specifically, $D_{i}$ is defined as:
\begin{equation}
   D_{i} = \left \{d_{1},d_{2},...,d_{H}\right \} ,2\le i\le T,
\end{equation}
where $H$ is the total number of edges with a weight change from $G_{i-1}$ to $G_{i}$, and each changed edge $d_{k}$ is:
\begin{equation}
   d_{k} = (u,w^{\prime}_k,v) ,1\le k\le H,
\end{equation}
where $w^{\prime}_k$ is the edge weight change of the $k$-th edge  between $G_{i-1}$ and $G_{i}$ and can be positive or negative.
A positive edge weight change indicates an increase of edge weight, while a negative one indicates a decrease of edge weight.


Like existing dynamic weighted graph visualizations~\cite{yi2010timematrix,bach2014visualizing}, we focus on the changes of edge weight, one of the important features of dynamic weighted graphs.
\newmod{
It is possible that some nodes may also appear or disappear in a dynamic weighted graph at some timeslices, which can be indicated by the appearance or disappearance of their associated edges. 
The dotted node in Figure~\ref{fig:teaser}a shows an example of node disappearance.}

%% file: source/4-diffSeer.tex
\section{DiffSeer}

\begin{figure*}
  \centering
  \mbox{} \hfill
  \includegraphics[width=0.98\linewidth]{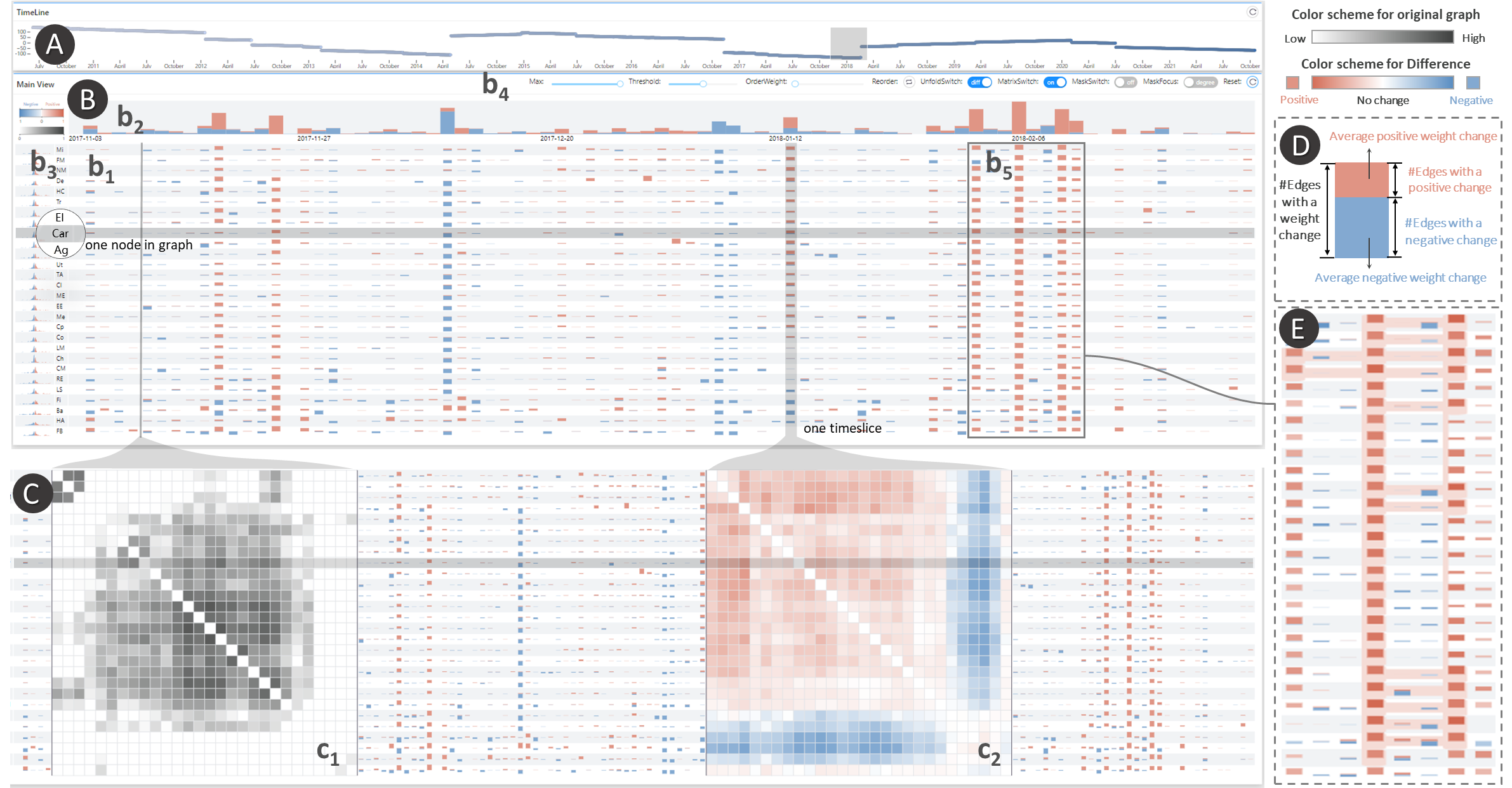}
  \hfill
  \mbox{}
  \caption{\mod{The \techName{} interface consists of 
  (A) a timeline view and (B) a nested matrix design including ($b_{1}$) the nested overview and detail matrices, ($b_{2}$) a stacked bar chart, ($b_{3}$) some area charts representing the overview of changes in nodes weights, and ($b_{4}$) a toolbar to provide some necessary interactions. The rows and columns in the nested matrix represent nodes and timeslices respectively.}
  (C) shows the nested matrix when an original detail matrix ($c_{1}$) and a difference detail matrix ($c_{2}$) are unfolded in the overview matrix. (D) is the explanation of each cell in the overview matrix. (E) shows the difference mask attached to ($b_{5}$) in the overview matrix.}
   \label{fig:system}
\end{figure*}

\mod{
We propose~\techName{}, a novel difference-based approach for dynamic weighted graph visualization.
Its core component is a nested matrix design (Figure~\ref{fig:system}B) to provide both an overview and fine-grained details of edge weight differences.
A difference mask (Figure~\ref{fig:system}E) can be interactively enabled to emphasize the significant changes, and stacked bar charts (Figure~\ref{fig:system}$b_{2}$) and area charts (Figure~\ref{fig:system}$b_{3}$) to overview the edge weight changes at each timeslice and associated edge weight change distribution of individual node respectively.
}
Besides the nested matrix design,
we also present a timeline view (Figure~\ref{fig:system}A) to enable a temporal summary of the original dynamic graph. It is achieved by projecting the graph structures at each timeslice into one dimension and the offset on the vertical axis can be a rough indication of the changing intensity~\cite{van2015reducing}, providing a quick insight into the entire period.
A prototype system of \techName{} is available at \textcolor{blue}{\url{https://diffseer.github.io/}}.

\subsection{Nested Matrix Design} \label{sec:nested}
This section introduces the visual design and the node reordering strategy of the nested matrix design in detail.

\textbf{\textit{Nested overview and detail matrices.}}
The nested matrix design has two core parts: an \textit{overview matrix} (Figure~\ref{fig:system}$b_1$) shows the graph differences between adjacent timeslices over a period of time to inform users of the temporal evolution of dynamic graphs, and \textit{detail matrices} (Figure~\ref{fig:system}$c_1$,$c_2$) display the details of graph difference or original graph structure at individual timeslices when a user double-clicks a timeslice of the \textit{overview matrix}.


\textit{Overview matrix.}
As shown in Equation \ref{diff}, graph difference can also be regarded as a type of graph.
\mod{
To effectively visualize the overall graph differences along time within the limited space,
we aggregate edge weight differences into their connected nodes in the overview matrix, which is an effective way of aggregation for graph visualization~\cite{bach2014visualizing}.
The rows of the overview matrix (Figure~\ref{fig:system}$b_{1}$) represent graph nodes, the columns correspond to different timeslices, and each cell intuitively shows the weight changes of all edges connected to one node in the corresponding timeslice.
For each cell (Figure~\ref{fig:system}D), we use 
a stacked-bar glyph to encode the number of edges in the \textit{graph difference}, i.e., the number of edges with a weight change between two adjacent timeslices in the original dynamic weighted graph.
The height of the upper or bottom bars of a glyph encodes the number of edges with a positive or negative weight change, respectively.
The color scheme of bars refers to average weight changes of all the edges connected to a node in the \textit{graph difference}, and we use a diverging red-to-blue color scheme to encode it.}

\textit{Detail matrix.}
To check the details of the differences or the context in which the differences occurred, we allow users to interactively unfold the detail matrices of individual timeslices within the overview matrix by double-clicking the corresponding timeslices, as shown in Figure~\ref{fig:system}C.
We use the matrix layout to visualize graph details because the detail matrices can share the same node order with the overview matrix to help users maintain analysis continuity.
The detail matrix has two types: the \textit{difference detail matrix} (Figure~\ref{fig:system}$c_{2}$) and the \textit{original graph detail matrix} (Figure~\ref{fig:system}$c_{1}$).
The difference detail matrix displays the specific difference between adjacent timeslices, where the rows and columns both represent nodes, and each cell represents a weight change of an edge.
\mod{
Similar to the overview matrix, we also use a diverging red-to-blue diverging color scheme here.
The original graph detail matrix
visualizes the graph structure
in an individual timeslice,
allowing users to inspect the exact graph structure details.
A sequential gray-scale color scheme is used to encode the original edge weights.}

\begin{figure}
  \centering
  \includegraphics[width=0.95\linewidth]{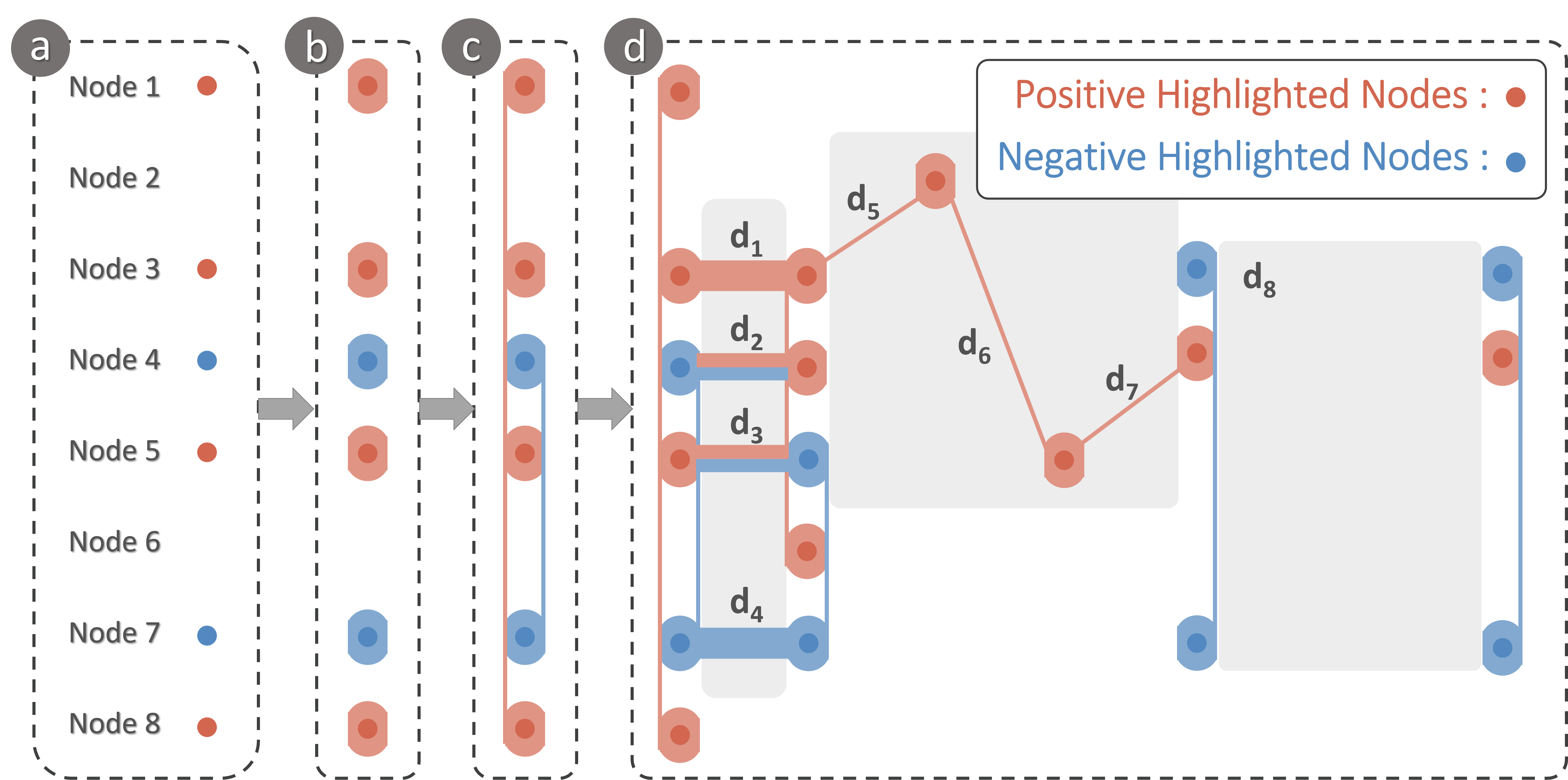}
    \setlength{\abovecaptionskip}{0.2cm}
    \setlength{\belowcaptionskip}{-0.2cm}
  \caption{\mod{The drawing procedure of \textit{difference mask}: (a) select nodes with significant changes, (b) highlight nodes, (c) connect highlighted nodes in one timeslice, (d) connect highlighted nodes cross timeslices.}}
  \label{fig:mask}
\end{figure}

\begin{figure}
  \centering
  \includegraphics[width=1\linewidth]{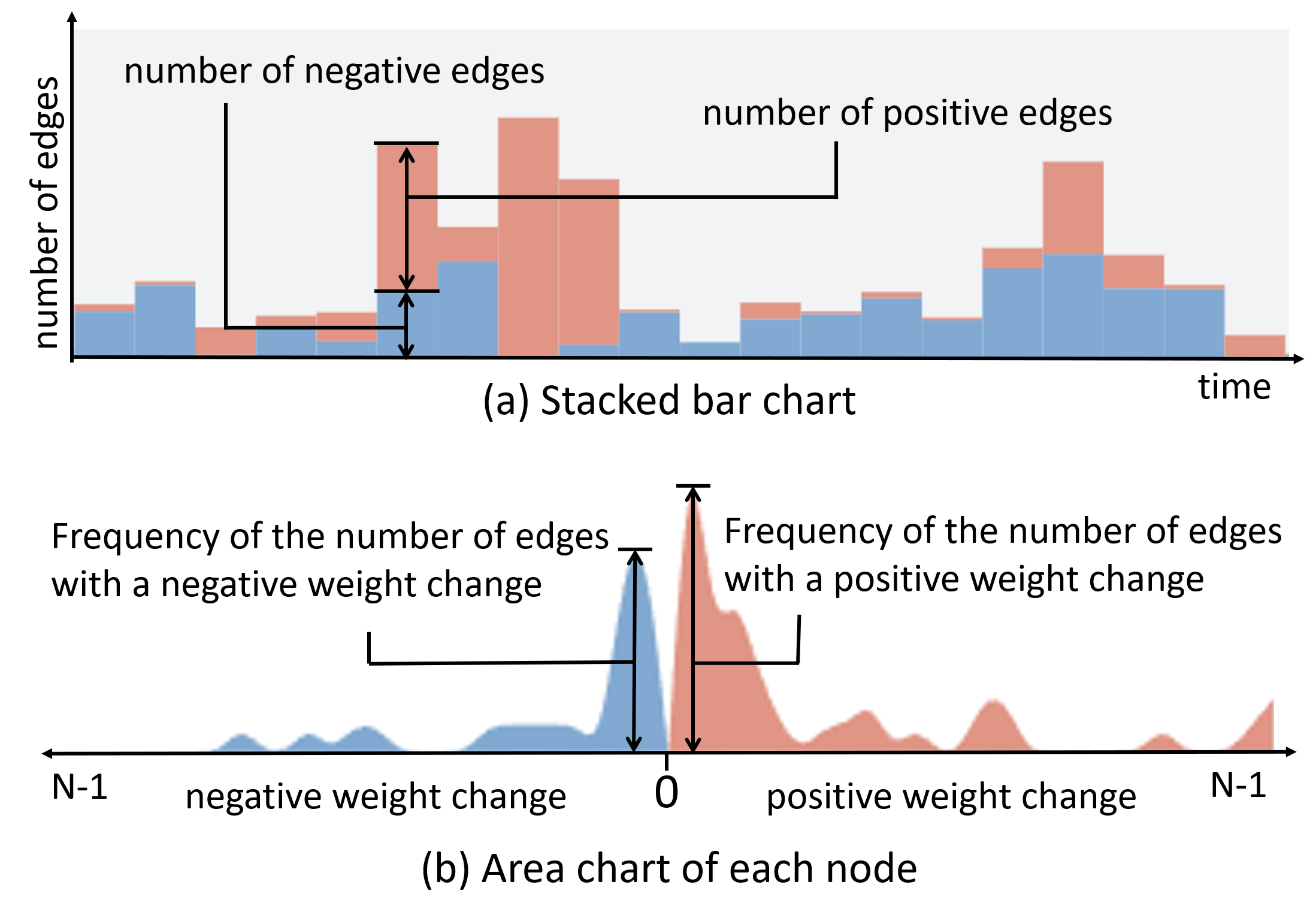}
    \setlength{\abovecaptionskip}{-0.4cm}

  \caption{\mod{
           Visual encoding of stacked bar chart (a) and area chart (b) of each node. N is the number of nodes in the graph.}}
    \label{fig:linkview}
\end{figure}

\textbf{\textit{Difference mask.}}
\mod{
To further facilitate an easy exploration of dynamic weighted graphs,
we propose a difference mask (Figure~\ref{fig:system}$E$) to emphasize the temporal patterns of graph differences (e.g., outliers or repeated changes~\cite{van2015reducing}).
Specifically, we highlight the nodes with a significant edge weight change above a user-defined threshold by using red and blue rounded rectangles (Figure~\ref{fig:mask}b). Here a significant edge weight change refers to either the average weight changes of edges connected to a node or the total number of edges with a weight change.
Then, we use paths to connect nodes with similar changes within and between timeslices, respectively.
}
\mod{For one timeslice, we use vertical paths to join the highlighted nodes of the same color in a column, red for positive and blue for negative (Figure~\ref{fig:mask}c), indicating that these nodes have similar significant changes.
To emphasize the temporal patterns, paths are also drawn between the columns with at least one highlighted node. 
Some of the paths are horizontal connecting the same highlighted nodes of two columns to denote that similar changes appear repeatedly, of which color depends on whether the two nodes are positive or negative (Figure~\ref{fig:mask}$d_{1-4}$).
In addition, when the number of consecutive columns with no highlighted nodes exceeds a predefined threshold, no path is drawn between them because this may imply a stable period (Figure~\ref{fig:mask}$d_{8}$).}

\textbf{\textit{Stacked bar chart and area chart.}}
\mod{
To provide the overall context, we add
a stacked bar chart and some area charts to the nested matrix design.}
The stacked bar chart (Figure~\ref{fig:linkview}a) presents the distribution of the changed edges' number over time, which shares the same timeline as the overview matrix, and the height of the red part encodes the number of the positive edges while the blue part encodes the negative edges.
The area charts (Figure~\ref{fig:linkview}b) present the distribution of the changed edges’ number of each node during the selected period, which can imply which nodes always have many edges with a weight change.

\subsection{Node Ordering Strategy}
\begin{figure}
  \centering
  \includegraphics[width=1\linewidth]{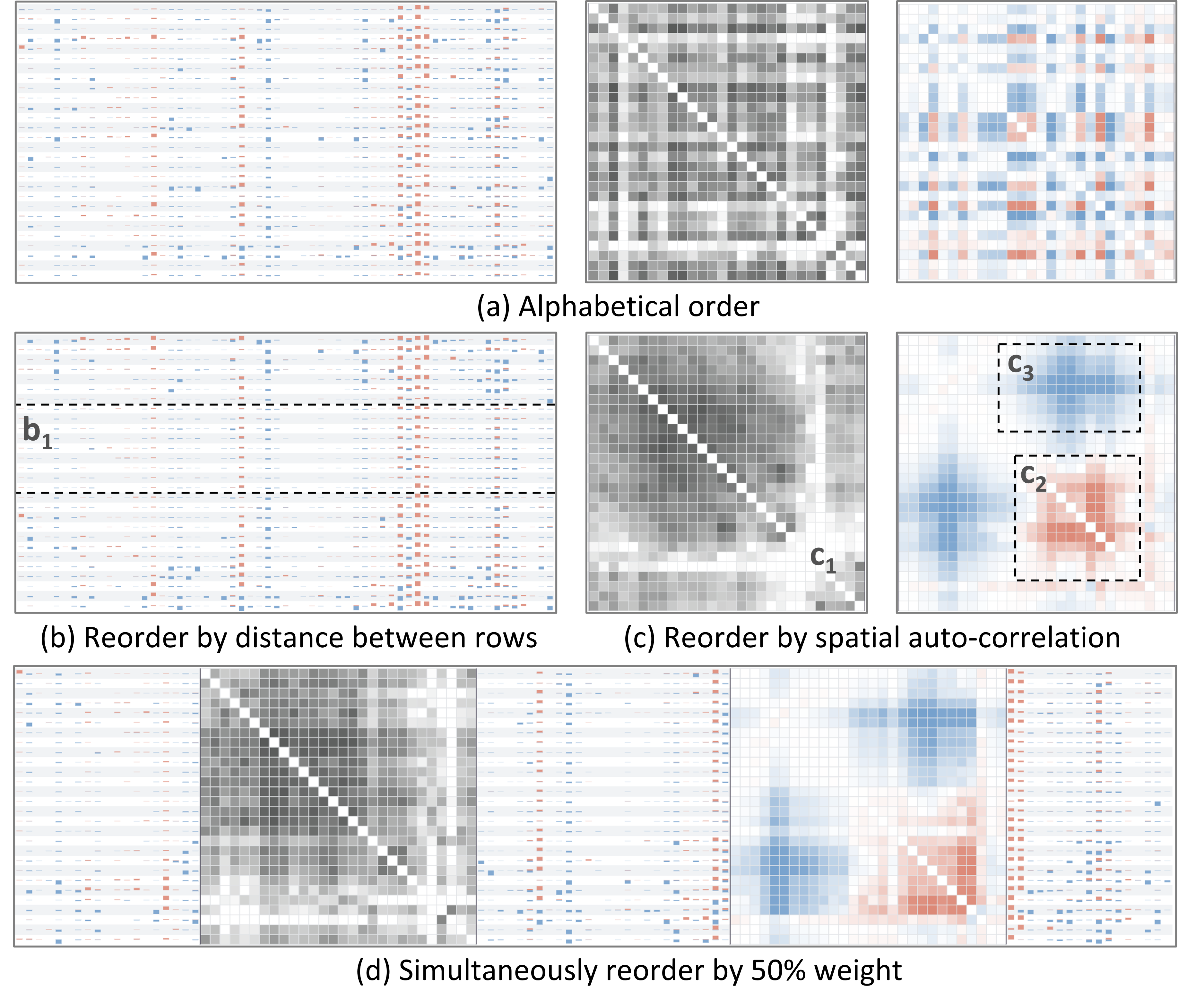}
  \caption{ The figures showcase the nested matrix design with different node orders to show the effectiveness of our node reordering strategy.
  (a) order by alphabetical. (b) order by the distance between rows, mainly designed for the overview Matrix. (c) order by spatial auto-correlation, mainly designed for the detail matrix. (d) order by considering both overview matrix and detail matrix equally. }
          \label{fig:reorder}
\end{figure}

The node order of a matrix-based design can directly determine the visualization effect~\cite{valdivia2019analyzing}.
In nested matrix design, overview matrix and detail matrices share the same node order, so a proper node order is of vital importance.
Figure~\ref{fig:reorder}a shows the overview matrix and detail matrices in an alphabetical node order.
We can discover some patterns in the overview matrix, such as the timeslices with significant changes, but it is hard to get deep insights like some nodes with similar patterns.
As for detail matrices, it is hard to figure out any interesting patterns beyond the value of each cell, such as clusters and subgroups.

To this end, we propose a node reordering strategy, which can simultaneously provide a single order that works well for both overview matrix and unfolded detail matrices.
Users can adjust the reordering strategy's priority for specific analysis needs.
Figure~\ref{fig:reorder}b,\ref{fig:reorder}c shows the overview matrix and detail matrices after node reordering, respectively.
Figure~\ref{fig:reorder}d shows the nested matrix after node reordering by treating the detail matrix and the overview matrix equally.
Compared to Figure~\ref{fig:reorder}a, each of the ordered matrices can provide some additional insights, such as some nodes with similar changing patterns (Figure~\ref{fig:reorder}$b_{1}$), two nodes that have little relationship to other nodes (Figure~\ref{fig:reorder}$c_{1}$), and a group of nodes that have stronger connections within them (Figure~\ref{fig:reorder}$c_{2}$) but weaker connections with other nodes (Figure~\ref{fig:reorder}$c_{3}$). 
Further, the node reordering by equally considering both detail matrix and overview matrix can also preserve their own visual patterns as much as possible (Figure~\ref{fig:reorder}d).

Most existing methods reordered the matrix by calculating the inner-rows distance of the matrix and then solving it as a traveling salesman problem (TSP)~\cite{behrisch2016matrix}.
Our strategy follows a similar framework, but the difference is that we need to find an appropriate order for different types of matrices simultaneously.
\mod{
Specifically, we augmented the reordering method using Moran's I~\cite{van2021simultaneous}, which is designed for unweighted and undirected dynamic graphs, to work for weighted dynamic graphs.}
Maximizing Moran's $I$ can be translated into minimizing the similarity $I_{s}$ between two adjacent rows:
\begin{equation}
  I= {\textstyle \sum_{a=1}^{n-1}}I_{s}(M,\rho\left ( a \right ),  \rho\left ( a+1 \right ) ) ,
\end{equation}
where $\rho\left ( a \right )$ denotes the a-th row of matrix $M$ ($n$ rows) in any node order and $ I_{s}\in \left [ -\frac{1}{n-1}, \frac{1}{n-1} \right ] $. 
We normalize the distance between two rows as $D_{detail}$.
For the overview matrix, since the spatial auto-correlation index can not work in the condition that the rows and columns are not symmetric, we only consider the distance between the two rows based on similarity. 
The inner-row distances are calculated with Manhattan distance as $D_{overview}$.

\mod{After getting the distances between rows of different types of matrices, the final node order should be calculated based on the user-defined weight to satisfy the user's preference.
We normalize the distance matrices of both the detail matrices and the overview matrix to the same scale and then
set an adjustable weight parameter $  \alpha \in \left [ 0, 1 \right ] $ to capture the user's preference, and the final inter-row distance $ D $ is defined as:
\begin{equation}
    D = \alpha D_{overview} + \left ( 1-\alpha \right ) D_{detail},
\end{equation}
which is used as the input of the leaf order algorithm~\cite{fekete2015reorder}.
It can calculate the final node order by constructing a hierarchical clustering on node rows according to the distance matrix.
}

When $\alpha$ is adjusted to 1, the reordering algorithm only focuses on the overview matrix (Figure~\ref{fig:reorder}b).
On the contrary, when $\alpha$ is 0, nodes are reordered only based on Detail Matrices (Figure~\ref{fig:reorder}c).
When $\alpha$ is set to a value between 0 and 1, such as 0.5, an overall suitable order may be obtained as shown in Figure~\ref{fig:reorder}d.

\subsection{Interactions}\label{sec:interaction}

\techName{} enables rich interactions to allow users to smoothly analyze and explore dynamic weighted graphs.
Users can brush the time range of their interest in the timeline view and explore the details in the nested matrix view.
\techName{} provides users with the flexibility of configuring the visual design, such as setting the reordering weight and the difference mask threshold. 
More details of interactions can be found in the prototype system.

%% file: source/5-usecases.tex
\section{Case Study}
We conducted two case studies to demonstrate the effectiveness of \techName{} on real-world dynamic weighted graph datasets, which are collected from the domains of finance and social network.
\mod{Two expert users (U1, U2) were involved in the case studies, and they have attended our user interviews and learned how to use \techName{}.
They were asked to use~\techName{} to explore dynamic weighted graphs. Their exploration procedures and the corresponding findings were recorded.
We used two dynamic weighted graphs whose details are as follows:
}


\textit{Sector Index Correlation Network (SICN).}
We collected time-series records of 28
sector indices in the Chinese stock market\footnote{\url{https://www.joinquant.com}}.
\mod{It consists of the stock trading records
from June 3, 2010 to October 18, 2021 (2,761 timeslices), which covers the time period of the COVID-19 outbreak.} We calculated the Pearson correlation between any two indices each day to model the association between different indices~\cite{simon2018hunting}.
The sector indices are regarded as the nodes, and the correlations between each pair of sector indices are regarded as the edge weight.

\textit{Rugby Team Tweet Network (RTTN).}
This dataset contains more than 3,000 tweets between 12 teams in the Guinness Pro12 competition from 2014 to 2015~\cite{simonetto2017drawing}. Each team is regarded as a node, and the edge weight denotes the number of tweets between two Rugby teams.
There are 12 nodes and 329 timeslices in total.

\subsection{Case 1: Impact of the COVID-19 on China Stock Market}

\begin{figure}
  \centering
  \includegraphics[width=1\linewidth]{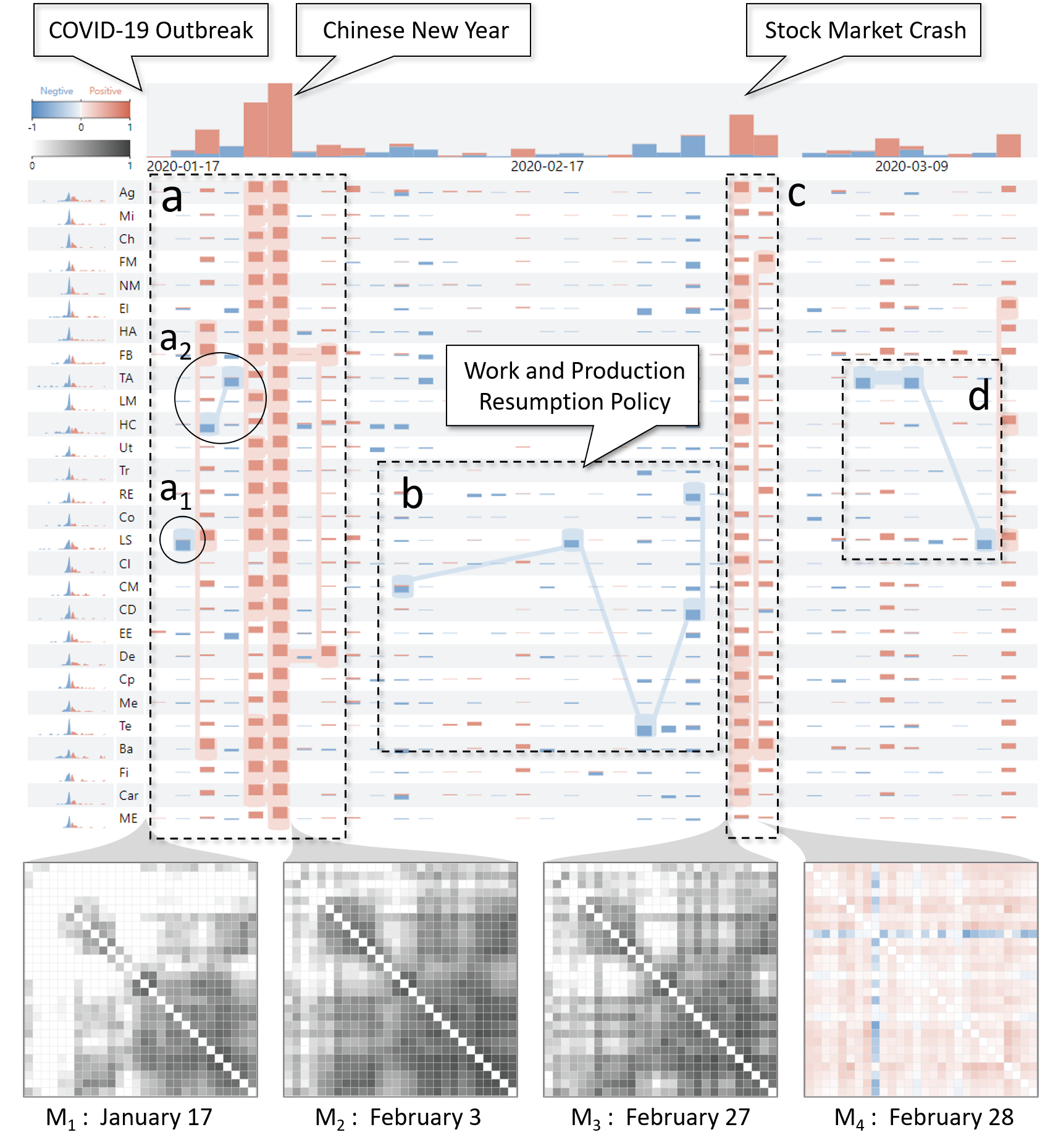}
  \caption{\label{fig:case1-2}
           Overview matrix of the sector index correlation network dataset after the COVID-19 outbreak. (a) and (c) show a significant increase in the sector correlations. (b) and (d) show that several sectors have weakened their connection with others.
           }
\end{figure}

U1 is a financial expert, and he was quite interested in
the impact of COVID-19 on the correlation between different sector indices in China stock markets. So he brushed the period after the COVID-19 outbreak (i.e., from January 17, 2020, to March 13, 2020), as shown in Figure~\ref{fig:case1-2}.
\mod{
From the overview matrix, U1 gained a quick understanding of the overall temporal evolution of the sector correlation network across 35 trading days.}
Specifically, he quickly noticed two periods with a significant increased in the sector correlations: the Chinese New Year period of 2020 and around early March, as indicated by the dense red bars in Figure~\ref{fig:case1-2}a and Figure~\ref{fig:case1-2}c.

\textbf{\textit{Chinese New Year of 2020.}}
\mod{
With~\techName{}, U1 easily noticed that there were significant increases of edge weights for many nodes before Chinese New Year (Figure~\ref{fig:case1-2}a), indicating a global correlation increase among different sectors.
At the beginning of the outbreak of COVID-19, it was observed that Node \textit{LS} (Leisure Service) first experienced a significant negative change (Figure~\ref{fig:case1-2}$a_{1}$), followed by the blue difference mask between Node \textit{HC} (Health Care) and Node \textit{TA} (Textile \& Apparel) (Figure~\ref{fig:case1-2}$a_{2}$).
It makes sense because Leisure Service (\textit{LS}), a crowd-intensive sector, was first affected by the COVID-19 pandemic, and (\textit{HC}) and (\textit{TA}) declined in correlation with other sectors due to the shortage of treatment, masks, and other protective resources.}
U1 unfolded two original graph detail matrices to check the context at the beginning and end of this period.
He found several clusters before the outbreak (Figure~\ref{fig:case1-2}$M_{1}$), but almost all the nodes were connected after the Chinese New Year (Figure~\ref{fig:case1-2}$M_{2}$), indicating that these sectors were responding in a similar manner to the outbreak.

\textbf{\textit{Work and production resumption.}}
U1 further identified several nodes connected by the
blue difference
mask in Figure~\ref{fig:case1-2}b.
\mod{
Except for these nodes,
the edge weights of all the other sector nodes
did not change very much during this period.
According to U1, the reason for this pattern was a succession of national policies aimed at these changed sectors, i.e., these nodes all had similar changing patterns after the introduction of the policy to resume work and production. 
Observing the detail matrix after the resumption (Figure~\ref{fig:case1-2}$M_{3}$), he found that the color of these rows was lighter than those in $M_{2}$ and confirmed that the work and production resumption did make some sectors less relevant to others.}

\textbf{\textit{Stock market crash.}}
Next, U1 wanted to analyze the significant correlation increase in Figure~\ref{fig:case1-2}c.
He unfolded the difference detail matrix
(Figure~\ref{fig:case1-2}$M_{4}$) by double-clicking
the corresponding timeslice.
U1 found only an apparent blue column (row) of Node \textit{TA} appearing in a wide range of red and noticed a negative difference mask on Node \textit{TA} and Node \textit{LS} after that day (Figure~\ref{fig:case1-2}$d$).
\mod{
U1 mentioned that the global increase of edge weight may result from the stock market crash, and the interesting pattern on Node \textit{TA} and Node \textit{LS} is owing to the policy support that kept these sectors isolated from the negative influence of the whole market.}

By using \techName{}, U1 concluded that the COVID-19 pandemic has had a great impact on the industry association network, with significant fluctuations in closely related sectors. 
U1 said that gaining these insights would have taken much longer without using \techName{}.


\subsection{Case 2: Tweet Interactions between Rugby Teams}

\begin{figure}[htb]
  \centering
  \includegraphics[width=0.95\linewidth]{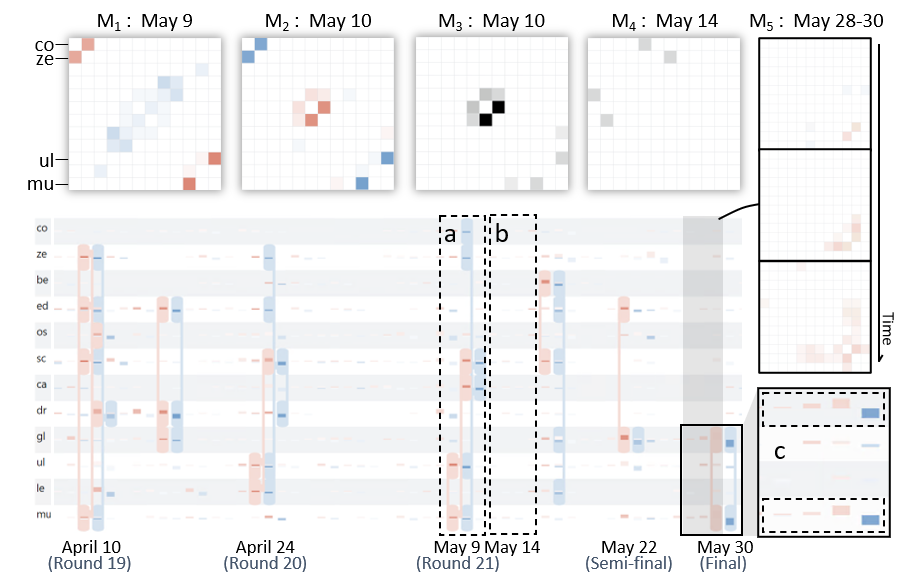}
    \setlength{\abovecaptionskip}{0.cm}
    \setlength{\belowcaptionskip}{-0.4cm}
  \caption{\label{fig:case2-1}
           Dynamic evolution of the rugby team tweet network dataset from April to June. (a) and (b) show some significant changes occurred during the game, but few changes occurred after the game. (c) shows the changed edges' number of two nodes increased continuously before the finals.}
\end{figure}

U2 has a background in social network analysis and is curious about the evolution of the Rugby team tweet network.
With \techName{}, U2 found interesting Tweet interaction patterns between Rugby teams from April 2015 to June 2015 (Figure~\ref{fig:case2-1}).

\textbf{\textit{Similar highlighted shapes.}}
\mod{
In the overview matrix, U2 found that some similar highlighted shapes were always separated by several columns with almost no changes,
where positive edge weights (red rectangles) often appear before negative edge weights (blue rectangles), as shown in Figure~\ref{fig:case2-1}.
It indicates recursive repeats of abrupt increase followed by abrupt decrease of edge weights.
To identify the reason behind it, U2 checked the schedule of the Rugby competitions for these teams~\footnote{\url{https://en.wikipedia.org/wiki/2014-15_Pro12}}.}
He found that these patterns are extremely relevant to the rounds of the competitions, i.e.,
there is a significant increase of the tweets between different Rugby teams on the start date of each round of competition, which would cease after the competition.
To further check the interaction details,
he unfolded two difference detail matrices in one of the highlighted shapes (Figure~\ref{fig:case2-1}a) and saw that two pairs of nodes were dark red in Figure~\ref{fig:case2-1}$M_{1}$ but dark blue in Figure~\ref{fig:case2-1}$M_{2}$.
\mod{
It turned out that they were exactly the four teams participated in the two matches on that day, indicating the abrupt increase or decrease of interaction mainly come from the teams involved in the competition.
Then, he compared the original graph detail matrices of a game day (Figure~\ref{fig:case2-1}$M_{3}$) and a day without a competition (Figure~\ref{fig:case2-1}$M_{4}$).
He confirmed that the stable period (e.g., Figure~\ref{fig:case2-1}b) was due to almost no tweets on days without games.}

\textbf{\textit{A pre-game hype.}}
\mod{
In Figure~\ref{fig:case2-1}c, U2 found that the red bars of the two nodes kept increasing from May 27 to May 30, indicating an increasing interaction between the two teams and other teams. but they suddenly switched to blue bars on June 1st.
By reordering nodes and comparing difference detail matrices, U2 found an obvious spread pattern of the two nodes, as shown in Figure~\ref{fig:case2-1}$M_{5}$, which shows that more and more teams started to have tweet interactions with these two teams.}
According to the schedule, May 30 was exactly the final round date between the two teams, so U2 inferred that this pattern probably results from the pre-game hype, which is also why the connections between teams decreased rapidly after the finals.

Overall, with \techName{}, U2 can figure out the deep correlation between the competition schedule and the evolution of the tweet interactions between Rugby teams.

%% file: source/6-expertinterview.tex
\section{User Interview}
\label{sec-user-interview}
We conducted semi-structured interviews with 12 expert users who work on network data analysis and visualization to evaluate the effectiveness and usability of \techName{}.

\subsection{Participants and Apparatus}
We invited 12 target users (U1-U12) to participate in the interviews.
U1 and U2 are researchers from financial and internet companies, and U3-U12 are students from several universities (two doctors and eight masters).
The participants (eight males and four females, aged 25 to 32, with normal vision) are engaged in network data analysis or visualization.
Four (U1-U4) have more than five years of research experience, and others have at least one year.
Due to the COVID-19 pandemic, our interviews with 7 participants were conducted online via Zoom, where they used their own computers to access \techName{} deployed on a Cloud server.
The interviews with the remaining participants were conducted offline on a desktop with a 23.8-inch 1920 x 1080 monitor.

\subsection{Datasets and Tasks}
\mod{
The SICN and RTTN datasets were used in our user interviews.
\mod{A small period of the RTTN dataset was used for the tutorial, while the other period of the RTTN dataset and the whole SICN dataset was used for the exploration by participants.
During the interviews, the participants were asked to complete the following five tasks to fully explore the dataset with \techName{}.}}

T1. Find the changes occurring over time.

T2. Describe the type of changes.

T3. Find repeated changes in this period.

T4. Find the nodes with similar change patterns over time.

T5. Find the nodes with similar patterns in one timeslice.

T1 and T2 are designed to test whether the nested matrix design can accurately represent differences and help participants quickly identify graph structures that have changed over time.
T3 is used to guide participants to discover some temporal patterns of differences with \techName{} since the repetition of changes is one of the major temporal patterns~\cite{van2015reducing}.
T4 and T5 aim to evaluate the usefulness of our node reordering strategy.
Since these tasks are subjective for users, we only use them to provide guidance but do not discuss the accuracy and time consumption.


\subsection{Procedure}
During the interview, we first introduced \techName{} to the participants.
Then, we went through an example usage scenario on the RTTN dataset to show how to use \techName{} to explore a dynamic weighted graph.
Then, the participants were allowed to explore the RTTN dataset freely to make themselves familiar with \techName{}.
The tutorial above lasted about 20 minutes.
\mod{
After that, they were invited to use \techName{} to analyze either the RTTN dataset of another time period or the SICN dataset 
and finish the above tasks.
Participants could freely explore the dataset
until they felt that the tasks had been well finished.
Their comments and suggestions were recorded.
After that,
we further invited them to finish a post-study questionnaire, including the set of questions in Table \ref{tab:table 2}.
Nine closed-ended questions (Q1-Q9) were designed to evaluate the usability (Q1-Q4) and effectiveness (Q5-Q9) of \techName{}, and two open-ended questions (Q10, Q11) were used to gather suggestions from participants. }
Overall, the whole interview lasted about 60 minutes for each participant.

\begin{table}
\caption{The questionnaire for user interview. Q1-Q9 are closed-ended questions (Q1-Q4 for usability and Q5-Q9 for effectiveness). Q10-Q11 are open-ended questions.
}\label{tab:table 2}
\centering
\begin{tabular}{c|p{6cm}}
    \toprule
    ID & \multicolumn{1}{c}{Questions} \\ \midrule
    Q1 &Is \techName{} easy or hard to learn? \\ 
    Q2 &Is \techName{} easy or hard to use?  \\ 
    Q3 &Is the visual design easy or hard to understand? \\
    Q4 &Is the interaction helpful or not in the analysis process?  \\  \midrule
    Q5 &Is it easy or hard to understand differences across timeslices with the overview matrix?  \\ 
    Q6 &Is it easy or hard to identify nodes with significant changes in the difference mask? \\
    Q7 &Is it easy or hard to check connection changes in each timeslice by unfolding the detail matrices?\\ 
    Q8 &Is the node reordering strategy helpful or not to enhance the visual pattern (like clustering) of the nested matrix?   \\ 
    Q9 &Overall, is the proposed method helpful or not for identifying the temporal evolution of dynamic weighted graphs?\\ \midrule
    Q10 &What are the advantages of the proposed method? \\ 
    Q11 &Which part of the method can be improved? How?      \\ \bottomrule
\end{tabular}   
\end{table}

\subsection{Results}
\mod{
Participants always got satisfactory answers for all tasks within 15 minutes, and we summarized the participants' standard of tasks as follows.
Firstly, participants found the changes over time by observing the blue/red cells in the overview matrix (T1). They checked both overview and detail matrices and described the changes by the changed edge number and weight (T2).
For T3, they found some repeated changes linked by difference mask. For T4 and T5, they reordered nodes and found nodes with similar changes gathering together.
}

Figure~\ref{interview} summarizes the participants' responses to our post-study questionnaire.
Overall, both the usability and effectiveness of \techName{} are highly rated by participants.
They agree that the \techName{} is convenient and efficient for exploring the temporal evolution of dynamic weighted graphs.
However, two of them thought that the visual design might not be friendly for a novice to understand. 
The detailed comments from participants are summarized as follows:

 \begin{figure}
 \centering   
 \includegraphics[width=1\linewidth]{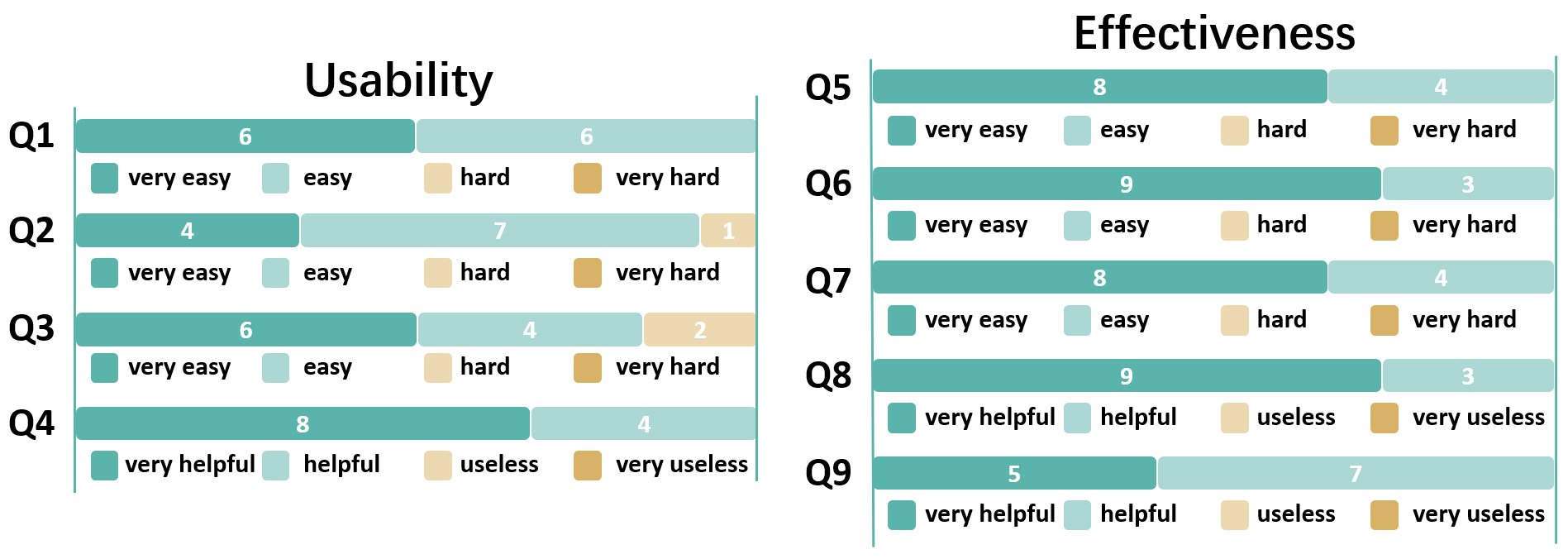}      
 \setlength{\abovecaptionskip}{-0.5cm}  
 \setlength{\belowcaptionskip}{-0.2cm}
 \caption{\label{interview} User interview results. Q1-Q4 is for usability, and Q5-Q9 is for effectiveness.} 
 \end{figure}
 
\textbf{\textit{Usability.}}
All participants confirmed that they could easily learn and use \techName{}, and the visual designs are intuitive. 
\mod{
Some of them have used small-multiples or animation to explore dynamic weighted graphs before, and they confirmed that they can find changes of edge weight more quickly and accurately with \techName{}.
U1 mentioned that ``\textit{Since matrices are very common in graph analysis, I can easily understand the nested matrix design.}"
U5 commented, ``\textit{It takes me some time to learn the meaning of Difference Mask, but it is quite useful after understanding it.}"
In terms of the interaction, the participants appreciated the existing interactions of \techName{}, which were confirmed to be beneficial and could satisfy the analysis requirements.}

\textbf{\textit{Effectiveness.}}
\mod{
Overall, all participants praised the effectiveness of \techName{} in exploring the dynamic weighted graph evolution.
They pointed out that the nested matrix design can help them quickly identify dynamic graph evolution characteristics, such as abrupt significant edge weight increases and recurrent edge weight changes. They said that it is more effective than the existing methods like small multiples.
}
U3 pointed out that ``\textit{With the node reordering strategy, the detail matrix can be more helpful to confirm the graph details within one timeslice.
But as for the weight of node reordering, I almost only chose 0 or 1 to focus on one matrix type since a median value is hard to interpret.}"
\mod{
However, U2 said that the adjustable weight feature helped him find a balance to analyze multiple matrices simultaneously.}



%% file: source/7-discussion.tex
\section{Discussion}
This section further discusses the generalizability, time range, and node number of \techName{}.

\textbf{\textit{Generalizability.}}
We have shown the generalizability of \techName{} by two case studies on real-world datasets with different characteristics, and the results can prove that \techName{} is useful for different types of dynamic weighted graphs that need to be analyzed for changes.
Our method can also support the dynamic unweighted graphs by considering the presence or absence of edges as an edge weight of 0 or 1.
Although \techName{} does not focus on the directed dynamic graphs, we can just encode one of the out-degree or in-degree in the overview matrix and reorder nodes with inner-row similarity instead of the auto-correlation index, then \techName{} can still work as expected.

\mod{
\textbf{\textit{Scalability.}}
\techName{} has good scalability in terms of the time range to be explored.
The overview matrix can help explore more than one thousand timeslices simultaneously since the column width can be narrow enough as long as users can identify it.
The time range can be further extended by brushing on the timeline view.
}


%% file: source/8-conclusion.tex
\section{Conclusion and Future Work}
We propose \techName{}, a dynamic weighted graph visualization approach, by explicitly visualizing the differences on edge weight between adjacent timeslices, which incorporates a novel nested matrix design, a new node reordering strategy, and a rich set of interactions to help users fully understand the temporal evolution of dynamic graphs.
For evaluation, we conduct two case studies on real-world datasets and in-depth interviews with 12 target users, and the results demonstrate that \techName{} is useful and effective in visualizing dynamic graphs.

\mod{
In future work, we plan to characterize differences from more perspectives, such as the removal or addition of nodes, to support broader analysis requirements.
It is also interesting to further explore how the proposed difference-based visualization approach can be extended to the visualization of other datasets, like high dimensional time-series data.
}

